\documentclass{IEEEtran}
\usepackage{cite}
\usepackage{amsmath,amssymb,amsfonts}
\usepackage{algorithmic}
\usepackage{graphicx}
\usepackage{bm}
\usepackage{subfigure}
\usepackage{caption}
\usepackage{textcomp}
\def\BibTeX{{\rm B\kern-.05em{\sc i\kern-.025em b}\kern-.08em
    T\kern-.1667em\lower.7ex\hbox{E}\kern-.125emX}}
\begin{document}
\title{Greedy Signal Space Recovery Algorithm with Overcomplete Dictionaries in Compressive Sensing}
\author{Jianchen Zhu, \IEEEmembership{Stduent member, IEEE}, Shengjie Zhao, \IEEEmembership{Senior member, IEEE}, Qingjiang Shi, \IEEEmembership{Member, IEEE}, Gonzalo R. Arce, \IEEEmembership{Fellow, IEEE}
\thanks{This work was supported in part by the National Basic Researh Program of China (973 program) under Grant 2014CB340404 and in part by the National Science Foundation of China under Grant 61471267.}
\thanks{J. Zhu is with the College of Electronics and Information Engineering, Tongji University, Shanghai 201804, China (e-mail: 1610486@tongji.edu.cn).}
\thanks{S. Zhao is with the Key Laboratory of Embedded System and Service Computing, Ministry of Education, Tongji University, Shanghai 201804, China (e-mail: shengjiezhao@tongji.edu.cn).}
\thanks{Q. Shi is with the School of Software Engineering, Tongji University, Shanghai 201804, China (e-mail: shiqj@tongji.edu.cn).}
\thanks{G. R. Arce is with the Department of Electrical and Computer Engineering, University of Delaware, Newark, DE 19716 USA (e-mail: arce@ee.udel.edu).}}

\maketitle

\begin{abstract}
Compressive Sensing (CS) is a new paradigm for the efficient acquisition of signals that have sparse representation in a certain domain. Traditionally, CS has provided numerous methods for signal recovery over an orthonormal basis. However, modern applications have sparked the emergence of related methods for signals not sparse in an orthonormal basis but in some arbitrary, perhaps highly overcomplete, dictionary, particularly due to their potential to generate different kinds of sparse representation of signals. To this end, we apply a signal space greedy method, which relies on the ability to optimally project a signal onto a small number of dictionary atoms, to address signal recovery in this setting. We describe a generalized variant of the iterative recovery algorithm called Signal space Subspace Pursuit (SSSP) for this more challenging setting. Here, using the Dictionary-Restricted Isometry Property (D-RIP) rather than classical RIP, we derive a low bound on the number of measurements required and then provide the proof of convergence for the algorithm. The algorithm in noisy and noise-free measurements has low computational complexity and provides high recovery accuracy. Simulation results show that the algorithm outperforms best compared with the existing recovery algorithms.
\end{abstract}


\section{Introduction}
\label{sec:introduction}
\IEEEPARstart{C}{ompressive} sensing\cite{candes2005decoding,donoho2006compressed} is a recently developed and fast growing field of research as a novel sampling paradigm. Suppose that $ x $ is a length-$ n $ signal. It is said to be $ k $-sparse (or compressible) if $ x $ can be well approximated using only $k \ll n$ coefficients under the some transform 
\begin{equation}\label{key}
x = \psi a,
\end{equation} 
where $\psi $ is the sparsifying basis and $ a $ is the coefficient vector that has at most $ k $ nonzero entries.

For a $ k $-sparse signal x, CS samples its $ m $ ($m < n$) random linear projections towards irrelative directions, which constitute measurements $ y $ with noisy perturbations $ e $. The process is simply described as 
\begin{equation}\label{key}
y = Ax + e,
\end{equation}
where $A \in {R^{m \times n}}$ represents the sensing matrix, and $ e $ is the acquisition noise. Since $m < n$ holds, problem (2) is ill-posed, and the perturbations $ e $ lead to unstable solutions. However, using the fact that $ x $ is sparse such that ${\left\| x \right\|_o} \le n$, it is possible recover $ x $ exactly via solving an ${l_1}$-minimization problem
\begin{equation}\label{eq}
\hat x = \arg \min {\left\| x \right\|_1}~~~s.t.{\left\| {y - Ax} \right\|_2} \le \varepsilon,
\end{equation}
where ${\left\| e \right\|_2} \le \varepsilon $ bounds the norm of the noise vector $ e $. Therefore, $ x $ can be exactly recovered by solving problem (3) provided the conditions on the Restricted Isometry property (RIP) are satisfied. 

$\bm{ Definition}$ $\bm{ 1 }$\cite{baraniuk2008simple}$\bm{ : }$ The sensing matrix $A \in {R^{m \times n}}$ is said to satisfy the k-order RIP if for any k-sparse (where ${\left\| x \right\|_0} \le k$) signal $x \in {R^n}$
\begin{equation}\label{eq}
	(1 - \delta )\left\| x \right\|{}_2^2 \le {\left\| {Ax} \right\|_2} \le (1 + \delta )\left\| x \right\|_2^2,
\end{equation}
where $\delta  \in [0,1]$. The infimum of $\delta $ , denoted by ${\delta _k}$,  is called the restricted isometry constant (RIC) of $ A $
\begin{equation}\label{eq}
	{\delta _k}: = \inf \{ \delta :(1 - \delta )\left\| x \right\|{}_2^2 \le {\left\| {Ax} \right\|_2} \le (1 + \delta )\left\| x \right\|_2^2\}. 
\end{equation}

Design of computationally efficient sparse signal recovery algorithms based on the RIP recovery conditions for the ${l_1}$-norm relaxation has extensively studied in previous works. Linear programming\cite{candes2005decoding} and other convex optimization algorithms\cite{randall2009sparse,candes2009exact} have been proposed to solve the problem (3). The most common approaches include the basis pursuit (BP), interior-point (IP)\cite{nesterov1994interior}, homotopy\cite{soussen2015homotopy}, and gradient projection for sparse representation (GPSR) algorithm\cite{blumensath2008gradient}. However, it has been shown that the sparse signal recovery problem can be solved with stability and uniform guarantees with polynomially bounded computation complexity.

As a result, Greedy Pursuit (GP)\cite{tropp2006algorithms} algorithms have also been widely studied. The predominant idea of GP algorithms is to estimate the nonzero elements of a coefficient vector iteratively. Matching Pursuit (MP)\cite{mallat1993matching} is the earliest greedy pursuit algorithm. The Orthogonal Matching Pursuit (OMP)\cite{tropp2007signal} is a well-known greedy pursuit algorithm as the improved version of MP. Several other advanced GP algorithms have been proposed, such as the Regularized OMP (ROMP)\cite{needell2010signal}, Compressive Sampling Matching Pursuit (CoSaMP)\cite{needell2009cosamp}, and Subspace Pursuit (SP)\cite{dai2009subspace}. Generally speaking, the GP algorithms have received considerable attention due to low computation complexity, high recovery accuracy, and simple implementation.

In some cases, the signal of interest is not itself sparse, but has a sparse representation in a overcomplete dictionary $ D $. Examples are found in a wide range of applications\cite{Donoho2001Sparse,Gilbert2005Applications,Rao1998Signal}, including image\cite{Cilingiroglu2015Dictionary}, audio\cite{Reyes2010Wavelet}, video compression\cite{Lin2006Video}, and source localization\cite{Malioutov2005A}. Proposals for sparse signal representations in an overcomplete dictionary include multiscale Gabor functiuons\cite{Qian1994Signal}, systems defined by algebraic codes\cite{Justesen2003Class}, wavelets and sinusoids\cite{Chen1998Atomic}, and multiscale windowed ridgelets\cite{Starck2002The}.

Numerous methods, both heuristic and theoretical, have been developed to support the benefits of such sparse signal representations: in theoretical neuroscience it has been argued that sparse signal representations in an overcomplete dictionary are necessary for use in biological vision systems\cite{Olshausen2009Learning}; in approximation theory, it has been demonstrated that approximation from overcomplete systems outperforms any known basis\cite{Nirmala2016A}; in signal processing, learned overcomplete dictionaries from a set of realizations of the data (training signals) are highly adapted to the given class of signals and therefore usually exhibit good representation performance\cite{Aharon2006}; and in image processing, the learned dictionaries have shown promising results in several recently published works on compression of facial images\cite{Ram2014Facial}, fingerprint images\cite{Shao2014Fingerprint}, geometry images of 3D face models\cite{Hou2014Expression}, synthetic aperture radar (SAR) images\cite{CHEN2012Multi}, and hyperspectral images\cite{İ2015Lossy}.

Consider the sparse signal representation based on atoms in a dictionary as columns in the matrix $D \in {R^{n \times d}}$. A sparse representation of the signal $x \in {R^n}$ can be thought of as a coefficient vector $a \in {R^d}$ in the dictionary $ D $. The representation is overdetermined if $d > n$. Hence, the representation of the signal in dictionary $ D $ is not unique, namely, there exists a variety of coefficient vectors that can be used to synthesize the signal . Furthermore, when the columns of the dictionary are high correlated and under a measurement $ y $, the matrix $ AD $ may not longer satisfy the condition of RIP. Hence, coefficient space methods, which aim at recovering the coefficient vectors, encounter a bottleneck due to the lack of orthogonality of the dictionary.

In this paper, we present a method based on the structure of the signal and the minimization of the gradient pursuit. Our main contribution is to develop a method, which we call the signal space  method\cite{giryes2015greedy,gu2016practical,davenport2013signal}. The advantage of the method is that it has the ability to optimally project the signal onto a small number of dictionary atoms under the true and correct support hypotheses at each iteration. By leveraging this technique, we present a novel GP algorithm called Signal Space Subspace Pursuit (SSSP). Furthermore, when the sensing matrix $ A $ satisfies the condition of D-RIP, we extend the algorithm for the accurate recovery on the sparse signal in an overcomplete dictionary and thus provide a proof convergence, using the D-RIP. In particular, rigorous guarantees for bound condition are derived showing that the algorithm recovers ideal sparse representation with the recovery error that grows at most proportionally to the noise level. Finally, the simulations demonstrate that the algorithm provides significant gains in the perfect recovery performance compared to that of the existing greedy algorithms as well as ${l_1}$-minimization algorithm via BP.  

The rest of the paper is organized as follows. We begin in Section II with a description of the mathematical model and its advantages. Section III develops our proposed algorithm and gives the convergence and recovery condition of the proposed algorithm in detail. Section IV provides some experimental results of the algorithm and its comparison with other existing CS recovery algorithms. Finally, we make the conclusion in Section V.

\section{System Model on Overcomplete Dictionaries}

\subsection{Compressive Sensing Model}
Suppose we have the compressive measurements $y \in {R^{m \times 1}}$ of an unknown sparse signal $x \in {R^{n \times 1}}$ given by
\begin{equation}\label{key}
y = Ax + e,
\end{equation}
where $A \in {R^{m \times n}}$ ($m \le n$) is the sensing matrix and $e \in {R^{m \times 1}}$ is the system noise. The sparsity condition on the signal $ x $ is that $x = Da$, for some coefficient vector $a \in {R^{d \times 1}}$ with ${\left\| a \right\|_0} \le k \le n$, yielding a $ k $-sparse representation of the signal $ x $ with respect to the Dictonary $D \in {R^{n \times d}}$ ($n \le d$). Thus, our task is to recover $ x $ based on $ y $, $ A $ and $ D $. Before we elaborate the algorithm, we first elaborate the considered signal sparsity model and the analysis vector $D * x$ as an assumption of "analysis sparsity" for $ x $ in the following section.

\subsection{Signal Sparisty Model}

\begin{figure}
	\centering
	\includegraphics[width=\columnwidth]{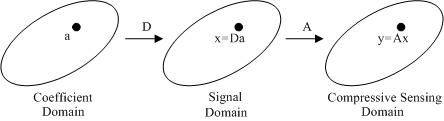}
	\caption{The compressive sensing process and its domains. This distinguish the domains in which the measurements, signals, and coefficients reside.}
	\label{fig:fig1}
\end{figure}

It is shown the model is discussed in the sparse-dictionary setting and recovery framework\cite{zhang2015optimal} for CS in Fig 1. Given a signal $ x $, let $ D $ be $n \times d$ matrix whose columns $D = \left[ {{D_1},{D_2},...,{D_n}} \right] \in {R^{n \times d}}$ from a Parseval frame for ${R^n}$, i.e.
\begin{equation}\label{eq}
	\begin{split}
		&x = \sum\limits_k {\left\langle {x,{D_k}} \right\rangle } {D_k} \\
		&\left\| x \right\|_2^2 = {\sum\limits_k {\left| {\left\langle {x,{D_k}} \right\rangle } \right|} ^2},\forall x \in {R^n}, \\
	\end{split}
\end{equation}
where $\left\langle {} \right\rangle $ denotes the standard Euclidean inner product. Notice that an overcomplete dictionary $D \in {R^{n \times d}}$ is a Parseval frame if $D{D^{\rm T}} = I$ and a coefficient vector $ a $ is compressible or k-sparse if ${\left\| a \right\|_0} \le k$. Then using a natural extension of the definition of RIP, the D-RIP is defined as follows.

$\bm{ Definition}$ $\bm{ 2 }$\cite{giryes2015greedy}$\bm{ : }$ Fix a overcomplete dictionary $D \in {R^{n \times d}}$. $ A $ is said to follow the restricted isometry property adapted to $ D $ (abbreviated D-RIP) with the constant ${\delta _k}$ if
\begin{equation}\label{eq}
\left( {1 - {\delta _k}} \right)\left\| {Da} \right\|_2^2 \le \left\| {ADa} \right\|_2^2 \le (1 + {\delta _k})\left\| {Da} \right\|_2^2
\end{equation}  
holds for all ${\left\| a \right\|_0} \le k$. 

Besides, for all k-sparse signals $ x $ in ${R^n}$, the required expected value of the constant ${\delta _k}$ can be calculated as
\begin{equation}\label{eq}
{\delta _k} = \mathop {\max }\limits_{T \subset \left\{ {1,...,d} \right\},{{\left\| T \right\|}_0} \le k} {\left\| {(AD)_T^{\rm T}{{(AD)}_T} - I} \right\|_2},
\end{equation}
where ${{{(AD)}_T}}$ is the submatrix of $ AD $ whose dimensions are $m \times \left| T \right|$ indexed by $\left| T \right|$ and ${\rm T}$ is the transposition operator.

Notice that when $ D $ is the identity, the definition of D-RIP reduces to the traditional definition of RIP. Numerous random matrices, such as Gaussian and Bernoulli matrices, satisfy the condition of D-RIP with high probability, which implies that the number of measurements required is on the order of $k\log (d/k)$. Such an assumption on $ A $ that modifications to (3) bound the recovery error for the ${l_1}$-analysis method implies that a signal $ x $ can be recovered from the noise-free and noisy measurements by solving the convex minimization problem such that
\begin{equation}\label{eq}
\hat x = \mathop {\arg \min }\limits_{\tilde x \in {R^n}} {\left\| {D*\tilde x} \right\|_1}~~~s.t.{\left\| {y - A\tilde x} \right\|_2} \le \mu, 
\end{equation}
where $\mu $ is the noise level with ${\left\| e \right\|_2} \le \mu $. The ${l_1}$-analysis method is based on the model assumption that for a signal $x = Da$ not only the coefficient vector $ a $, but also the analysis vector ${D * x}$ is sparse. Using the assummption tbat adding an additional factor ${\left\| {D * x - {{(D * x)}_k}} \right\|_1}/\sqrt k $, the upper bound of recovery error for the ${l_1}$-analysis method is give by $O({\left\| {D * x - {{(D * x)}_k}} \right\|_1}/\sqrt k  + {\left\| e \right\|_2})$, where ${{{(D * x)}_k}}$ is a best $ k $-sparse approximation of ${D * x}$. Notice that the term ${\left\| {D * x - {{(D * x)}_k}} \right\|_1}$ bounds the recovery error in this case. If the analysis vector ${D * x}$ has a suitable decay , the recovery error depends only on the noise level ${\left\| e \right\|_2}$ in the measurements. Under these assumptions, the convergence for the ${l_1}$-analysis method and the algorithm we design are both in proportion to ${\left\| e \right\|_2}$. Without loss of generality, the convergence for the algorithm similar to that for the ${l_1}$-analysis method (see Sections III-D, III-E and III-F below for details).

A weaker assumption on the convergence for the algorithm that all signals corresponding to the coefficient vector $ a $ implies that there exists a localization factor defined as follows.

$\bm{ Definition }$ $\bm{ 3 }$$\bm{ : }$ For a dictionary $D \in {R^{n \times d}}$ and a sparsity level $ k $, we define the localization factor as          
\begin{equation}\label{eq}
{\eta _{k,D}} = \eta \mathop  = \limits^{def} \mathop {\sup }\limits_{{{\left\| {Da} \right\|}_2} = 1,{{\left\| x \right\|}_0} \le k} {{{{\left\| {D*Da} \right\|}_1}} \over {\sqrt k }}.
\end{equation}

The localization factor can be viewed as a measure of how sparse the objective in problem (13) is. Notice that if $ D $ is orthonormal, then $\eta  = 1$ (proof: See appendix A), and $\eta $ increases with the redundancy in $ D $.

\section{Signal Space Subspace Pursuit Algorithm}
Firstly, we propose the algorithm based on GP algorithms in the sparse-dictionary framework. Secondly, using the condition of DIP, we provide a guarantee for the minimum number of measurements required. Thirdly, we derive a bound that theoretically provides a sufficient condition for exactly signal recovery to demonstrate probable performance of the algorithm.

\subsection{Algorithm Design}
An overview of the algorithm is introduced first. Then, the flow of the algorithm is given and several key steps are analyzed. Finally, the advantages of the algorithm are discussed in detail.

Before introducing the iterative algorithm for sparse signal recovery, the following notation will be used in the formulation of the recovery algorithm.

$\bm{ Notation }$$\bm{ : }$ Suppose that $ A $ is an $m \times n$ matrix. Suppose that we observe a set of noisy measurements of the form $y = Ax + e$. If there exists an index set (support set) $\Lambda  \subset \{ 1,2,...,d\} $, we let ${D_\Lambda }$ denote its submatrix ${D_\Lambda } \in {R^{n \times \left| \Lambda  \right|}}$ with columns index by $\left| \Lambda  \right|$, and we let $R({D_\Lambda })$ represent the column span of ${D_\Lambda }$. 

$\bm{ Remark }$$\bm{ : }$ Note that algorithm requires some knowledge about the sparsity level $ k $, and there are some effective approaches to approximate the parameter. One alternative approach is to conduct empirical studies with all sparsity levels and select the level which minimizes ${\left\| {y - AD\tilde a} \right\|_2}$.   

As it will be shown, the most remarkable novelty of the algorithm is that the signal can be recovered exactly in an overcomplete dictionary. This novelty makes the algorithm more general and improves the selection of effective atoms. The main steps of the algorithm are summarized below. 

\begin{table}[htbp]
	\centering
	\caption{Signal Space Subspace Pursuit Algorithm}
	\label{label1}
	\begin{tabular}{p{2.9in}}
		\hline
		$\bm{ Algorithm }$: Signal Space Subspace Pursuit                                                                                                   \\
		\hline
		\begin{tabular}[c]{@{}l@{}}$\bm{ Input }$: \\ Sensing matrix $ A $; Dictionary $ D $; measurements $ y $; \\ sparsity level $ k $; stopping criterion $\varepsilon $ \end{tabular}            \\ \hline
		$\bm{ Initialization }$: \\ Iteration time $l = 0$; support estimation $I = \emptyset $; initial residual ${r^0} = y$; initial approximate ${x^0} = 0$                                                                                                                   \\ \hline
		$\bm{ while }$ halting criterion is not satisfied $\bm{ do }$                                                                                                                                 \\ 
		\begin{tabular}[c]{@{}l@{}}$\bm{ S1 }$: Find the index $\Omega  = {S_D}(u,3k)$ with $u = A * r$ \\ \{index corresponding to the largest magnitude entries in \\ the product\}.
		\end{tabular} \\
		$\bm{ S2 }$: Find the support estimation $T = \Omega  \cup I$                                                                                                                                                 \\
		$\bm{ S3 }$: Calculate the signal estimation: \\ $\tilde x = D\tilde a = D(\arg {\min _{{a_T}}}{\left\| {y - AD\tilde a} \right\|_2}~~s.t.{a_{{T^C}}} = 0)$                                                      \\
		$\bm{ S4 }$: Shrink the index $I = {S_D}(\tilde x,k)$  =\{index corresponding to the   largest magnitude entries in estimated $\tilde x$\}                                                                                    \\
		$\bm{ S5 }$: Calculate the new signal estimation: ${x^{l + 1}} = {P_I}\tilde x$
		\\
		$\bm{ S6 }$: Calculate the new residual: ${r^{l + 1}} = y - A{x^{l + 1}}$                                                                                                                \\
		$\bm{ end }$ $\bm{ while }$: $l = MaxIter$ or ${\left\| {{x^{l + 1}} - {x^l}} \right\|_2}/{\left\| {{x^l}} \right\|_2} \le \varepsilon $ is satisfied                                                                                                                               \\ \hline
		\begin{tabular}[c]{@{}l@{}}$\bm{ Output }$: \\ Signal estimation $\hat x = {x^{l + 1}} = SSSP(A,D,y,k)$ 
		\end{tabular}                                                   \\ \hline
	\end{tabular}
\end{table}

Before calculating the product, the column vectors of the sensing matrix $A = [{A_1},...,{A_n}]$ should be normalized firstly in practice. The most relevant column of $ A $ for  residual error ${r^l}$ is select to minimize the next residual error ${r^{l + 1}}$.

\begin{figure}
	\centering
	\includegraphics[width=0.5\columnwidth]{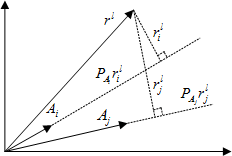}
	\caption{Projection of the residual error.}
	\label{fig:fig2}
\end{figure}

As can be seen in Fig. 2, although the product $\left| {\left\langle {{r^l},{A_i}} \right\rangle } \right|$ is bigger than $\left| {\left\langle {{r^l},{A_j}} \right\rangle } \right|$, the length of ${P_{{A_j}}}r_j^l$ is smaller than that of ${P_{{A_i}}}r_i^l$, thus the residual error $r_i^l$ is smaller than $r_j^l$. Therefore the one that has the smallest residual error or biggest projection length is selected. The projection length onto the column vector ${A_i}$ can be denoted as
\begin{equation}\label{eq}
	p = \left| {{{\left\langle {{r^l},{A_i}} \right\rangle } \over {{{\left\| {{A_i}} \right\|}_2}}}} \right| = \left| {\left\langle {{r^l},{{{A_i}} \over {{{\left\| {{A_i}} \right\|}_2}}}} \right\rangle } \right|,
\end{equation}  
where ${\left\| {{A_i}} \right\|_2}$ denotes the Euclidean length of  vector ${A_i}$, and ${{{A_i}} \over {{{\left\| {{A_i}} \right\|}_2}}}$ is the normalized  ${\left\| {{A_i}} \right\|_2}$. So in practice, the sensing matrix $ A $ should be normalized before calculating the product. 

Analogous to the classical GP algorithms, the most fundamental step is to calculate the observation $ u $ during each iteration, as shown in step 1 in the algorithm procedure. This common step occupies most part of the calculation in all GP algorithms even in the case where $ D $ is an overcomplete dictionary. Besides, the algorithm very similar to this appear in the analysis of the initialization procedure of these GP algorithms. The initial estimation matrix is the matrix ${x^0} = 0$ and thus the initial residual is the matrix of input measurements ${r^0} = y$. When an initial observation $u = A * {r^0}$ as the proxy for the support estimation $T$ is required. ${T^0} = \Omega  \cup {I^0}$ where ${\Omega ^0} = {S_D}(u,3k)$ is a subroutine identifying the index set of the rows of $ u $ with the $ 3k $ largest row-${\left\| l \right\|_2}$-norms. In iteration $ l+1 $, the algorithm update the previous estimation ${x^l}$ by taking a step of calculating the residual ${r^l}$ in the steepest descent direction $A * {r^l}$. A new support estimation, ${T^{l + 1}}$, is then obtained by merging previous supports estimation ${\Omega ^l}$ and ${I^l}$ such that ${T^{l + 1}} = {\Omega ^l} \cup {I^l}$. Finally, after $ l+1 $ iteration, a new support estimation ${I^{l + 1}}$ is obtained by taking a step of sharinking the index ${I^{l + 1}} = {S_D}(\tilde x,k)$. The algorithm updates the new solution ${x^{l + 1}}$ that minimizes the residual error ${\left\| {y - AD{a_{{I^{l + 1}}}}} \right\|_2}$ when restricted on ${I^{l + 1}}$, and calculate new residual ${r^{l + 1}} = y - A{x^{l + 1}}$. The choice of stopping criteria plays an important role for the algorithm, and the stopping criteria (such as normalized relative error ${\left\| {{x^{l + 1}} - {x^l}} \right\|_2}/{\left\| {{x^l}} \right\|_2} \le \varepsilon $) applies to the experiments are outlined in Section V.

Next, recall that some key steps in the classical GP algorithms. The GP algorithms identify nonzero entries of the support of the signal per iteration. Given a support estimation $ T $ and a constant number of iterations $ l+1 $, once the least square solution is obtained based on the corresponding support estimation such that 
\begin{equation}\label{key}
\tilde x = D (A_{{T^{l + 1}}}^\dag y) = D({(A_{{T^{l + 1}}}^{\rm T}{A_{{T^{l + 1}}}})^{ - 1}}A_{{T^{l + 1}}}^{\rm T}y).
\end{equation}
where $A_{{T^{l + 1}}}^\dag $ is the pseudoinverse of ${A_{{T^{l + 1}}}}$. If the accurate support estimation ${A_{{T ^{l + 1}}}}$ is provided, then $y = Ax = {A_{{T ^{l + 1}}}}{a_{{T ^{l + 1}}}}$ with $\det ({A_{{T ^{l + 1}}}}) \ne 0$, and so combining this result in (13) yields $x = \tilde x$. These steps are trivial and can be performed by simple thresholding of the entries of the coefficient vector in the case where $ D $ is orthonormal, i.e, $D = I$. Thus, our task in sparse-dictionary signal recovery is to correctly identify the support estimation $ T $. The algorithm we design solve this problem by iteratively identifying like columns, performing a projection which, given a general vector, find the closest $ k $-sparse vector, and then deciding which columns of $ A $ to choose. In the representation case (when $ D $ is an overcomplete dictionary ), a simple hard thresholding is replaced with an appropriate operator that takes a  candidate signal and finds the best $ k $-sparse representation of a vector $ z $. Towards this end, in the signal space, we define
\begin{equation}\label{eq}
	{\Lambda _{opt}}(z,k) = \mathop {\arg \min }\limits_{\Lambda :{{\left\| \Lambda  \right\|}_0} \le k} {\left\| {x - {P_\Lambda }z} \right\|_2},
\end{equation} 
where ${P_\Lambda }$ denotes the projection onto the span of the columns of $ D $ indexed by $\Lambda $. This problem is itself reminiscent of the conventional CS problem; one wants to recover a sparse representation from an underdetermined linear system\cite{davenport2013signal}. Thus, we make a conclusion that it is an NP-hard problem in general. Therefore, we allow for near-optimal projection to be used in the algorithm, writing ${S_D}(x,k)$ to denote the k-sparse approximation to $ x $ in $ D $. To denote the k-sparse approximation to ${S_D}$, the algorithm is surprisingly able to exactly recover $ x $.

In general, consider that (14) seems to be NP hard because it requires examining all $ k $  possible combinations of the columns in an overcomplete dictionary. To overcome this difficulty, an approximation is needed. For this we instead look for a near-optimal projection scheme, as used in our algorithm.  It has been shown that as long as the near optimal projection is good enough, namely,
\begin{equation}\label{eq}
	\begin{split}
		&{\left\| {{P_{{S_D}(x,k)}}x - x} \right\|_2} \le {c_1}{\left\| {{P_\Lambda }x - x} \right\|_2} \\
		&\left\| {{P_{{S_D}(x,k)}}x} \right\| \ge {c_2}{\left\| {{P_\Lambda }x} \right\|_2} \\
	\end{split}
\end{equation}              
for all $ x $ and suitable constant ${c_1}$ and ${c_2}$ (where ${P_\Lambda }$ denotes the optimal projection), then the algorithm provides accurate recovery of the signal. Although such projections for a well behaved dictionary $ D $ exists, however, such projections are not known to exist when the dictionaries are highly redundant. Interestingly, empirical studies using classical GP algorithms for such projections showed that the algorithm using these projections still yields exact recovery in such setting. 

\subsection{Bound for the Number of Measurements}
Above all, setting aside the question of how to design the dictionary $ D $, we address the problem of designing the sensing matrix $ A $. In the previous works, it can be shown that $ x $ can be  stably recovered from the compressive measurements $ y $ satisfying the classical condition of RIP with a small constant ${\delta _k} \in (0,1)$. However, numerous signals in pratice are compressible in the overcomplete dictionary. Due to the effect of redundancy, the recovery error ${\left\| {x - \hat x} \right\|_2}$ in signal space can be significantly smaller or larger than the recovery error ${\left\| {a - \hat a} \right\|_2}$ in coefficient space.

We now turn to the case where x is compressible in an overcomplete dictionary $ D $. Specifically, the matrix $ A $ satisfies the condition of D-RIP of order $ k $ if there exists a constant ${\delta _k} \in (0,1)$ such that 
\begin{equation}\label{eq}
	\sqrt {1 - {\delta _k}}  \le {{{{\left\| {ADa} \right\|}_2}} \over {{{\left\| {Da} \right\|}_2}}} \le \sqrt {1 + {\delta _k}} 
\end{equation}
holds for all $ a $ satisfying ${\left\| a \right\|_0} \le k$. the condition of D-RIP ensures norm preservation of all signals having a sparse representation $x = Da$. Thus, the condition of RIP is considered to be a stronger requirement when $ D $ is an overcomplete dictionary.  

There are numerous methods to design matrices that satisfy the condition of D-RIP. To the best of our knowledge, the  commonly used random matrices satisfy the condition of RIP or D-RIP with high probability. 

More speciffic, we consider matrices constructed as follows: we generate a matrix $A \in {R^{m \times n}}$ by selecting the entries $A[M,N]$ as independent and identically distributed random variables. We impose two conditions on the random distribution. First, we require that the distribution is centered and normalized such that $A[M,N]\mathop  \sim \limits^{idd} N(0,{1 \over m})$. Second, we require that the random variable $\left\| {Ax} \right\|_2^2$ in \cite{baraniuk2008simple} has expected value $\left\| x \right\|_2^2$; that is,
\begin{equation}\label{eq}
	E\left( {\left\| {Ax} \right\|_2^2} \right) = \left\| x \right\|_2^2.
\end{equation}

Generally speaking, any distribution, which includes the Gaussian and uniform distribution, with bounded support is subgaussian.

The key property of subgaussian random variables that will be of used in this paper is that any matrix $A \in {R^{m \times n}}$ which for a fixed vector (signal) $ x $ satisfies
\begin{equation}\label{eq}
	\Pr \left( {\left\| {Ax} \right\|_2^2 - \left\| x \right\|_2^2 \ge \varepsilon \left\| x \right\|_2^2} \right) \le 4{e^{ - {c_0}(\varepsilon )m}}.
\end{equation} 

It implies that the matrix $ A $ will satisfy the condition of D-RIP with high probability as long as $ m $ is at least on the order of $k\log (d/k)$. From this, the probability is taken over all draws of $ A $ and the constant ${c_0}(\varepsilon )$ rely both on the particualr subgaussian distribution and the range of $\varepsilon $. Perhaps the most important for our purpose is the following lemma.

$\bm{ Lemma }$ $\bm{ 1 }$$\bm{ : }$ Let $ Δ ：\chi $ denote any k-dimensiinal of ${R^n}$. Fix $\delta ,\alpha  \in \left( {0,1} \right)$. Suppose that $ A $ is a $m \times n$ random matrix with i.i.d entries chosen from a distribution satisfying (18), we obtain the minimal number of measurements required for exact recovery
\begin{equation}\label{eq}
	m = O({{2k\log (42/\delta ) + \log (4/\alpha )} \over {{c_0}(\delta /\sqrt 2) }}),
\end{equation} 
then with probability exceeding $1 - \alpha $,
\begin{equation}\label{eq}
	\sqrt {1 - {\delta }} {\left\| x \right\|_2} \le {\left\| {Ax} \right\|_2} \le \sqrt {1 + {\delta }} {\left\| x \right\|_2}
\end{equation} 
for all $x \in Δ ：\chi $.

Proof: See Appendix B.

When $ D $ is an overcomplete dictionary, one can use Lemmma 1 to go beyond a single k-dimensional subspace to instead considering all possible subspace spanned by k columns of $ D $, thereby establishing the condition of D-RIP for $ A $. Then, we have the following Lemma.

$\bm{ Lemma }$ $\bm{ 2 }$$\bm{ : }$ Let D be an overcomplete dictionary whose dimension is $n \times d$ and fix $\delta ,\alpha  \in \left( {0,1} \right)$. we obtain the minimal number of measurements required for exact recovery
\begin{equation}\label{eq}
	m = O({{2k\log (42ed/\delta k + \log (4/\alpha )} \over {{c_0}(\delta /\sqrt 2 )}})
\end{equation}

The proof follows that of Appendix B.

with $ e $ denoting the base of the natural logarithm, then with probability $1 - \alpha $, $ A $ will satisfy the condition of D-RIP of order $ k $ with the constant $\delta $. 

As noted above, the random matrix approach is somewhat useful to help us solve signal recovery problems. In this paper, we will further focus on the random matrices in the development of our theory. 

\subsection{Bound for the Tail Energy}
In this section, we focus on the tail energy since it plays a important role in our analysis of the convergence of the algorithm (see Section III-E below for details). In particular, we give some useful expansions to demonstrate the bound condition of the tail energy.

$\bm{ Definition }$ $\bm{ 4 }$$\bm{ : }$ Suppose that $ x $ is a k-sparse signal in the overcomplete dictionary $ D $ domain and $ e $ is an additional noise (where ${\left\| e \right\|_2} \le \varepsilon $), then we have
\begin{equation}\label{eq}
	\tilde e = {\left\| {x - {x_k}} \right\|_2} + {{{{\left\| {x - {x_k}} \right\|}_1}} \over {\sqrt k }} + {\left\| e \right\|_2},	
\end{equation}  
where ${x_k}$ is the best approximation of $ x $. The algorithm makes significant progress at each iteration where the recovery error is large relative to the tail energy. In noisy case, the tail energy as the quantity measure the baseline recovery error.

Assume that $ p $ is a number in the interval $(0,1)$. Let the signal $ x $ is p-compressible with magnitude $ R $ when the components of  $ x $ such that $\left| {{x_1}} \right| \ge \left| {{x_2}} \right| \ge  \cdots  \ge \left| {{x_n}} \right|$ obey a a power law decay such that 
\begin{equation}\label{eq}
	\left| {{x_i}} \right| \le R*{i^{ - 1/p}},~~~\forall x = 1,2, \cdots n.
\end{equation}

According to (23), ${\left\| x \right\|_1} \le R*(1 + \log n)$ when $p = 1$ and p-compressible signal is almost sparse when $p \approx 0$. In general, the p-compressible signals apply to approximate sparse signals such that 
\begin{equation}\label{eq}
	\begin{split}
		&{\left\| {x - {x_k}} \right\|_1} \le {D_1}*R*{k^{1 - 1/p}}, \\
		&{\left\| {x - {x_k}} \right\|_2} \le {D_2}*R*{k^{1/2 - 1/p}}, \\
	\end{split}
\end{equation}
where the constants ${D_1} = {(1/p - 1)^{ - 1}}$ and ${D_2} = {(2/p - 1)^{ - 1/2}}$. Note that (24) provides upper bounds on the two different norms of the recovery error $\left\| {x - {x_k}} \right\|$. Combining this result with (22), the tail energy in a p-compressible signal is upper bounded by 
\begin{equation}\label{}
	\tilde e \le 2{D_1}*R*{k^{1/2 - 1/p}} + {\left\| e \right\|_2}.
\end{equation}

When the parameter $ p $ is enough small, the most term in the right hand (25) decays rapidly as the sparsity level $ k $ increases.

\subsection{Recovery of Approximately Sparse-Dictionary Signals from Incomplete Measurements}
Consider that signals have a sparse representation in an overcomplete dictionary $ D $, we theoretically provide a guarantee for exact recovery of sparse-dictionary signals. Analogously to the guarantees of GP algorithms, the proof relies on iteration invariant which indicates that the recovery error is  mostly determined by the number of iterations. 

Before stating the main result for the algorithm (Theorem 3), we first state the following Theorem .

$\bm{ Theorem }$ $\bm{ 2 }$ Assume that $ A $ satisfy the condition of D-RIP with the constant ${\delta _{4k}} < 0.1$. Let ${S_D}(x,k)$ be the near optimal projections in (15) and ${x^{l + 1}}$ be the approximation after $l + 1$ iterations. if $(1 + {c_1})(1 - {{{c_2}} \over {{{(1 + \beta )}^2}}}) < 1$, the upper bound of recovery error after $l + 1$ iterations is given by
\begin{equation}\label{key}
{\left\| {x - {x^{l + 1}}} \right\|_2} \le {\eta _1}{\left\| e \right\|_2},
\end{equation}
where $\beta $ is an arbitrary constant, and ${\eta _1}$ is a constant which depends on ${c_1}$, ${c_2}$ and $\beta $. Inspired by the precious work in the signal space setting, the conditions of Theorem 2 on the near optimal projections holds in cases where $ D $ is not unitary and especially in cases where $ D $ is highly overcomplete/redundant that can't satisfy the traditional condition of RIP. To the best of our knowledge, the classical GP algorithms are used to calculate the projections. Thus, we provide a stronger convergence for the algorithm even when the dictionary is highly overcomplete in the following Theorem. 

$\bm{ Theorem }$ $\bm{ 3 }$$\bm{ : }$ Let $ A $  be a sensing matrix satisfying the condition of D-RIP of order $4k$ or a coefficient vector $ a $ such that $x = Da$. Then, the signal estimation ${x^{l + 1}}$ after $l + 1$ iterations of the algorithm satisfies
\begin{equation}\label{eq}
	\begin{split}
		&{\left\| {x - {x^{l + 1}}} \right\|_2} \le {C_1}{\left\| {x - {x^l}} \right\|_2} + {C_2}{\left\| e \right\|_2} \\
		&{\rm with} \\
		&{C_1} = ((2 + {\lambda _1}){\delta _{4k}} + {\lambda _1})(2 + {\lambda _2})\sqrt {{{1 + {\delta _{4k}}} \over {1 - {\delta _{4k}}}}} \\
		&{C_2} = ({{(2 + {\lambda _2})((2 + {\lambda _1})(1 + {\delta _{4k}}) + 2)} \over {\sqrt {1 - {\delta _{4k}}} }}). \\
	\end{split}
\end{equation}

Proof: The proof follows that of Theorem II.1\cite{davenport2013signal}

Notice that constants ${C_1}$ and ${C_2}$ that depended on the isometry constant ${\delta _{4k}}$ and on the approximation parameters ${\lambda _1}$ and ${\lambda _2}$. Further, an immediate Lemma of Theorem 3 is the following.

$\bm{ Lemma }$ $\bm{ 3 }$$\bm{ : }$ Assume that the conditions of Theorem 3. Then after a constant number of iterations 
${l + 1} = \left\lceil {{{\log ({{\left\| x \right\|}_2}/{{\left\| e \right\|}_2})} \over {\log (1/{C_1})}}} \right\rceil $ it holds that
\begin{equation}\label{eq}
{\left\| {x - {x^{l + 1}}} \right\|_2} \le (1 + {{1 - {C_1}^{{l + 1} }} \over {1 - {C_1}}}){C_2}{\left\| e \right\|_2}.
\end{equation}

Proof: See Appendix C.  

Notice that Lemma 3 implies the results, Theorem 2, with ${\eta _1} = (1 + {{1 - C_1^{l + 1}} \over {1 - {C_1}}}){C_2}$. 

More specifically, through various combinations of ${c_1}$, ${c_1}$ and ${\delta _{4k}}$, Theorem 3 shows that ${C_1} < 1$ and the accuracy of the algorithm improves per iteration. Thus, we obtain ${C_1} \le 0.5$ and ${C_2} \le 7.5$ if ${\lambda _1} = {1 \over {10}}$, ${\lambda _2} = 1$, and ${\delta _{4k}} \le 0.1$. Applying the recursive nature of the Theorem 3, we have the following Lemma.

$\bm{ Lemma }$ $\bm{ 4 }$$\bm{ : }$ Suppose that the condition of Theorem 3 hold with the constant ${\delta _{4k}} \le 0.1$. For each iteration of the algorithm, the signal estimation ${x^l}$ after l-th iterations is k-sparse, and 
\begin{equation}\label{eq}
	\begin{split}
		&{\left\| {x - {x^{l + 1}}} \right\|_2} \le 0.5{\left\| {x - {x^l}} \right\|_2} + 7.5{\left\| e \right\|_2}. \\
		&{\rm Particularly,}      \\
		&{\left\| {x - {x^l}} \right\|_2} \le {2^{ - l}}{\left\| x \right\|_2} + 15{\left\| e \right\|_2}. \\
	\end{split}
\end{equation} 

Each iteration of the algorithm reduces the recovery error by a constant factor, while adding an additional noise component. By taking a sufficient number of iterations $ l $, the most term ${2^{ - l}}{\left\| x \right\|_2}$ can be made as small as possible, and ultimately the recovery error is proportional to the noise level in the noisy measurements. If the accurate ${S_D}$ is provided, the upper bound of recovery error in (29) also applied to those of commonly used results.                  

\subsection{Recovery of Approximately Arbitrary Signals from Incomplete Measurements}
As shown in the proof of Theorem 3, in the case where the signls have a sparse representation in $ D $, smaller values of ${c_1}$ and ${c_2}$ result in a more accurate recovery and it is possible to achieve accurate recovery as accurate as desired by choosing small enough of ${\left\| e \right\|_2}$. However, this is not the case that signals don't exactly have a sparse representation in $ D $, that is, if 
\begin{equation}\label{key}
y = A(D{a_k}) + A(x - D{a_k}) + e = A(D{a_k}) + \hat e.
\end{equation}

Notcie that the term $\hat e = A(x - D{a_k}) + e$ can be viewed as the noise in the noisy measurements of the k-sparse signal $D{a_k}$ with ${\left\| {{a_k}} \right\|_0} \le k$. In fact, the "new" noise $ \hat e $ bounds maximum achievable accuracy. For the sake of illustration, the condition of Lemma 4 still holds. Further, we state 2 Theorems and 2 Lemmas in this section, which can be considered as the extensions of Theorem 3 and its Lemma (Lemma 4) to this case.

First, we state the following lemma, which can be considered as a generalization to Lemma 4. 

$\bm{ Lemma }$ $\bm{ 5 }$$\bm{ : }$ For the general CS model $y = A(D{a_k}) + \hat e$ in (30), if ${\delta _{4k}} < 0.1$, the upper bound of recovery error is given by 
\begin{equation}\label{eq}
	\begin{split}
		&{\left\| {x - {x^{l + 1}}} \right\|_2} \le 0.5{\left\| {x - {x^l}} \right\|_2} + {\left\| {x - D{a_k}} \right\|_2} \\
		&+ 7.5{\left\| {A(x - D{a_k})} \right\|_2} + 7.5{\left\| e \right\|_2}. \\
		&{\rm Particularly,}  \\
		&{\left\| {x - {x^l}} \right\|_2} \le {2^{ - l}}{\left\| {D{a_k}} \right\|_2} + {\left\| {x - D{a_k}} \right\|_2} \\
		& + 15{\left\| {A(x - D{a_k})} \right\|_2} + 15{\left\| e \right\|_2}, \\
	\end{split}
\end{equation}
where ${{a_k}}$ is the best $ k $-sparse approximation of $ x $ with ${\left\| {{a_k}} \right\|_0} \le k$.

Proof: See appendix D.

Notice that the coefficient vector ${{a_k}}$ we choosed is used to minimize the upper bound of (31), which indicates that ${{a_k}}$ is still important for sparse signal recovery in measurements and specifically in noisy measurements. From Lemma 5, the term ${\left\| {A(x - D{a_k})} \right\|_2}$ can be used to prove the convergence of the algorithm when $ D $ is not unitary. The assumption that modifications to (31) implies that there exist an upper bound of the term ${\left\| {A(x - D{a_k})} \right\|_2}$ in the signal space as stated in the following Theorem. 

$\bm{ Theorem }$ $\bm{ 4 }$$\bm{ : }$ Suppose that $ A $ satisfies the upper bound of RIP with the constant ${\delta _{4k}} < 0.1$.Then, for any vector $x \in {R^n}$,
\begin{equation}\label{eq}
	{\left\| {Ax} \right\|_2} \le \sqrt {1 + {\delta _k}} ({\left\| x \right\|_2} + {1 \over {\sqrt k }}{\left\| x \right\|_1}).                        
\end{equation}

Proof: See appendix E.

Using this Theorem to bound the right hand of (31), we derive
\begin{equation}\label{eq}
\begin{split}
&{\left\| {x - {x^{l + 1}}} \right\|_2} \le 0.5{\left\| {{x} - {x^l}} \right\|_2} + 7.5{\left\| e \right\|_2} \\
& + (7.5\sqrt {1 + {\delta _k}}  + 1){\left\| {x - D{a_k}} \right\|_2} + {{7.5\sqrt {1 + {\delta _k}} } \over {\sqrt k }}{\left\| {x - D{a_k}} \right\|_1}. \\
&{\rm Particularly,}  \\
&{\left\| {x - {x^l}} \right\|_2} \le {2^{ - l}}{\left\| {D{a_k}} \right\|_2} + 15{\left\| e \right\|_2} \\
& + (15\sqrt {1 + {\delta _k}}  + 1){\left\| {x - D{a_k}} \right\|_2} + {{15\sqrt {1 + {\delta _k}} } \over {\sqrt k }}{\left\| {x - D{a_k}} \right\|_1}. \\
\end{split}
\end{equation} 

Denote 
\begin{equation}\label{key}
M(x): = \mathop {\inf }\limits_{{a_k}:{{\left\| {{a_k}} \right\|}_0} \le k} ({\left\| {x - D{a_k}} \right\|_2} + {1 \over {\sqrt k }}{\left\| {x - D{a_k}} \right\|_1}),
\end{equation}
is the model mismatch quantity (for any $x \in {R^n}$). Notice that (32) and (34) have very similar form even in the case where $ D $ is not a overcomplete dictionary. Combining this result in (33), we have 
\begin{equation}\label{key}
\begin{split}
&{\left\| {x - {x^{l + 1}}} \right\|_2} \le 0.5{\left\| {x - {x^l}} \right\|_2} + 7.5{\left\| e \right\|_2} + 8.5\sqrt {1 + {\delta _k}} M(x). \\
&{\rm Particularly,}  \\
&{\left\| {x - {x^l}} \right\|_2} \le {2^{ - l}}{\left\| {D{a_k}} \right\|_2} + 15{\left\| e \right\|_2} + 16\sqrt {1 + {\delta _k}} M(x). \\
\end{split}
\end{equation}

Notice that the quantity bounds the above recovery error. If the quantity we chose is enough large, then the signal is not a k-sparse signal or a compressible signal such that $x \ne D{a_k}$, which implies that signals still don't exactly have a sparse representation in $ D $.

Similarly, we derive an upper bound of recovery error which is nearly relative to the tail energy as stated in the following Theorem.

$\bm{ Theorem }$ $\bm{ 5 }$$\bm{ : }$ Let $ A $ be a sensing matrix satisfying the condition of D-RIP. Assume that ${\delta _{4k}} < 0.1$. Given the assumption that modifications to (33) for the general CS model, the upper bound of recovery error is given by
\begin{equation}\label{eq}
	\begin{split}
		&{\left\| {x - {x^{l + 1}}} \right\|_2} \le 0.5{\left\| {x - {x^l}} \right\|_2} + 10\tilde e. \\
		&{\rm Particularly,}  \\
		&{\left\| {x - {x^l}} \right\|_2} \le {2^{ - l}}{\left\| {D{a_k}} \right\|_2} + 20\tilde e. \\
	\end{split}
\end{equation} 

Proof: See appendix F.

After $ l+1 $ iterations, the term ${2^{ - l}}{\left\| {D{a_k}} \right\|_2}$ can be made enough small such that $\mathop {\lim }\limits_l {2^{ - l}}{\left\| {D{a_k}} \right\|_2} = 0$ and the recovery error depends only on the tail energy, which implies that the algorithm make significant progress per iteration in this case.

Recall that the term ${\left\| {A(x - D{a_k})} \right\|_2}$ bounds the recovery error in (31). The assumption that modifications to (31) implies that there exist an upper bound of the term ${\left\| {A(x - D{a_k})} \right\|_2}$ in the coefficient space as stated in the following Lemma. 

$\bm{ Lemma }$ $\bm{ 6 }$$\bm{ : }$ If $ AD $ satisfies the condition of D-RIP with the constant ${\delta _{4k}} < 0.1$, then using the extension of (32) yields
\begin{equation}\label{eq}
	\begin{split}
		&{\left\| {A(x - {x_k})} \right\|_2} = {\left\| {AD(a - {a_k})} \right\|_2} \\
		& \le \sqrt {1 + {\delta _k}} ({\left\| {a - {a_k}} \right\|_2} + {{{{\left\| {a - {a_k}} \right\|}_1}} \over {\sqrt k }}). \\
	\end{split}
\end{equation}

Using this Lemma that the term ${\left\| {AD(a - {a_k})} \right\|_2}$ bounds the recovery error in the coefficient space, we derive
\begin{equation}\label{eq}
	\begin{split}
		&{\left\| {x - {x^{l + 1}}} \right\|_2} \le 0.5{\left\| {x - {x^l}} \right\|_2} + 7.5{\left\| e \right\|_2} + {\left\| {x - D{a_k}} \right\|_2} \\
		& + 7.5\sqrt {1 + {\delta _k}} ({\left\| {a - {a_k}} \right\|_2} + {1 \over {\sqrt k }}{\left\| {a - {a_k}} \right\|_2}). \\
		&{\rm Particularly,}  \\
		&{\left\| {x - {x^l}} \right\|_2} \le {2^{ - l}}{\left\| {D{a_k}} \right\|_2} + 15{\left\| e \right\|_2} + {\left\| {x - D{a_k}} \right\|_2} \\
		& + 15\sqrt {1 + {\delta _k}} ({\left\| {a - {a_k}} \right\|_2} + {1 \over {\sqrt k }}{\left\| {a - {a_k}} \right\|_2}), \\
	\end{split}
\end{equation}
where ${a_k}$ is the best k-sparse approximation of $ a $. If ${a_k}$ we chose is arbitrarily compressible, then $a = {a_k}$, which implies that the upper bound of (38) is reasonably small.

\subsection{Computation Complexity of the Algorithm}
In this section, we further obtain the following result regarding the convergence speed of the algorithm.

Recall that $\hat x = {x^{l + 1}}$ as an output of the algorithm after $l + 1$ iterations. Given a postive parameter $\eta $, the algorithm produces a signal estimation $\hat x$  after at most $O(\log {\left\| x \right\|_2}/\eta )$ iterations such that
\begin{equation}\label{eq}
{\left\| {x - \hat x} \right\|_2} = O(\eta  + {\left\| e \right\|_2}) = O(\max \left\{ {\eta ,{{\left\| e \right\|}_2}} \right\}.
\end{equation} 

The cost of one iteration of the algorithm is dominated by the cost of steps 1 an 6 of the algorithm as Table I is presented. The first step is to obtain the proxy $u = A*r$  and the signal estimation $\tilde x$. The next step is to calculate the support approximation ${S_D}$ efficiently with the classical GP algorithms which includes OMP, ROMP, CoSaMP and SP are used to estimate ${S_D}$. The running time of these algorithms over an $n \times d$ dictionary $ D $ is $O(knd)$ or $O(nd)$. Therefore, the overall running time of these GP algorithms is $O(knd\log {\left\| x \right\|_2}/\eta )$ or $O(nd\log {\left\| x \right\|_2}/\eta )$. Notice that the dictionary $ D $ is overcomplete. For sparse signal recovery, these running time are in line with advanced bounds for the algorithm, which implies that the algorithm has linear convergence shown in Fig. 3. 

Interestingly, we now turn to the case where the number of measurements required is calculated through reducing the approximation recovery error if there exists the R-SNR. Thus, given a sparse-dictionary $ x $ with ${\left\| x \right\|_2} \le 2R$, the upper bound of SNR is given by  
\begin{equation}\label{eq}
\begin{split}
&R - SNR = 10\log ({{{{\left\| x \right\|}_2}} \over {{{\left\| {x - \hat x} \right\|}_2}}}) = 10\log ({{{{\left\| x \right\|}_2}} \over {{{\left\| {x - D\hat a} \right\|}_2}}}) \\
&\le 10\log ({{{{\left\| x \right\|}_2}} \over {{{\left\| {x - {x_k}} \right\|}_2}}}) = 10\log ({{{{\left\| x \right\|}_2}} \over {\tilde e}}) \\
&\le 10\log ({{2R} \over {2{D_1} \cdot R \cdot {k^{1/2 - 1/p}}}})= \log ({1 \over p} - 1) + ({1 \over p} - {1 \over 2})\log k,  \\
\end{split}
\end{equation}		
where ${\hat x}$ is the approximation of $ x $. The number of iterations required is $O(\log k)$. Therefore, if the fixed R-SNR can be guaranteed, the overall running time of the algorithm is $O(nd\log {\left\| x \right\|_2}/\eta  \cdot SNR)$ in this case, which further implies that the computation complexity of the algorithm is nearly linear in the signal length.

\begin{figure}
	\centering
	\includegraphics[width=9cm, height=6.5cm]{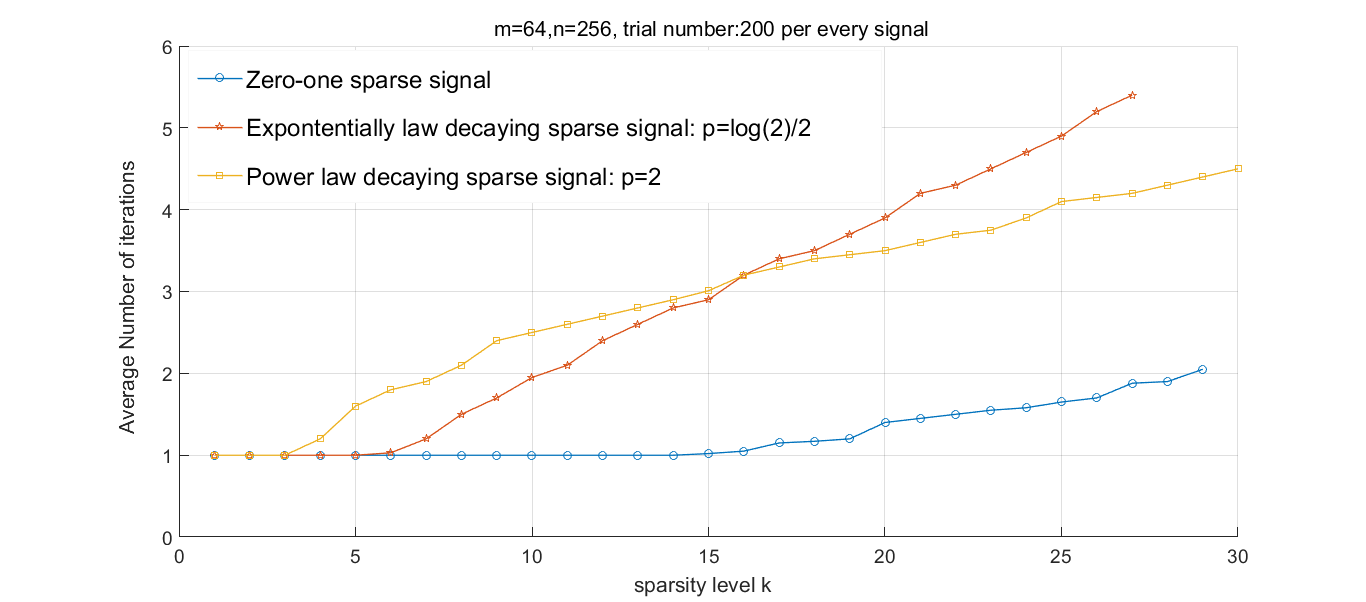}
	\caption{Convergence of the signal space subspace pursuit algorithm for three types of signals.}
	\label{fig:fig5}
\end{figure}

\section{Simulation Results}
\label{sec:guidelines}
This section tests the performance of the algorithm by conducting a wide range of numerical experiments. Above all, when the dictionary is an orthonormal basis, the influence of the sparsity level and the number of measurements required for its recovery performance is studied. Recovery performance analyses are further conducted by comparing the algorithm with some recently developed recovery algorithms being commonly used, including OMP, ROMP, CoSaMP, SP and LP. Next, note that if the dictionary is not an orthonormal basis, the main difficulty in implementing our algorithm is in calculating the projection of the signal onto a small number of dictionary atoms. To overcome this difficulty, we apply GP algorithms to approximate it. By conducting 1000 independent trials in all simulations, the algorithm using the signal space method can outperform the conventional algorithms.

In all experiments, simulated data are generated by taking the following steps:
\begin{enumerate}
	\item Generate a $ k $-sparse signal $ x $ of length $n = 256$ sparse in the dictionary $ D $ domain, i.e. $ x=Da $. Its coefficient vector $ a $ has $ k $ nonzero entries whose magnitudes are Gaussian distributed and locations are at uniformly random.
	\item Generate a sensing matrix $A \in {R^{m \times n}}$. Then entries of $ A $ are independently generated from Gaussian distribution.
	\item Compute the measurements by $y = Ax$ or $y = Ax+e$.     
\end{enumerate}

After the simulation data are generated, the above mentioned algorithms are used to recover a k-sparse signal $ x $ under the given $ A $ and $ y $.

To evaluate the estimation quality, two indices ${R_{rec}}$ and ${R_{res}}$ are commonly used. First, the recovery error ${R_{rec}}$ is defined by
\begin{equation}\label{eq}
	{R_{rec}}(x,\hat x) = {{{{\left\| {x - \hat x} \right\|}_2}} \over {{{\left\| x \right\|}_2}}} \le {\varepsilon _1}.
\end{equation}

We say that a signal $ x $ is exactly recovered when the signal estimation $\hat x$ satisfies ${\left\| {x - \hat x} \right\|_2} \le {10^{ - 4}}{\left\| x \right\|_2}$.

\subsection{Simulation Results on Analysing the Recovery Performance of the Algorithm under a Renormalized Orthogonal Dictionary}

In the first experiment, we evaluate the recovery performance of the algorithm and compare it with that of the five existing algorithms mentioned above. Note that the matrix $ D $ is an orthogonal but not a normalized basis. The signal $ x $ of the length $ n=256 $ is sparse in the Dictionary domain, i.e. $ x=Da $, where the dictionary $ D $ is the $256 \times 256$ matrix. Its coefficient vector $ a $ has $ k=10 $ nonzero entries whose magnitude are Gaussian distributed and locations is at uniformly random. We investigate the frequency of signal recovery as a function of the number of measurements. Simulation results are shown in Fig. 4. 

Recall that the problem (14) is NP-hard in our analytical framework because that it requires examining all $ k $ combinations of the columns of $ D $. To calculate ${\Lambda _{opt}}(x,k)$ with such a dictionary, we utilize the column norms of $ D $ to divide the $ k $ largest nonzero entries of the analysis vector $ D*x $ and their corresponding supports, which implies that sets $\Lambda $ equal to the positions of the $ k $ largest entries. 

As can be seen in Fig. 4(a), in the noise-free case, the algorithm improves the signal recovery frequency significantly compared to those of the five existing algorithms. For example, the algorithm recovers a k-sparse signal $ x $ with more than $90\% $ frequency up to the number of measurements $ m=55 $. Whereas, the LP-minimization algorithm with high computation complexity is able to recover $ x $ only up to the number of measurements $ m=60 $ under the same signal recovery frequency constraint. Moreover, as can be seen in Fig. 4(b), in the noisy case, the algorithm outperforms other algorithms by optimally approximating the supports. Further, as can be seen in Fig. 4, the algorithm utilize the matrix $ AD $ to recovery the coefficient vector $ a $ because of the nonremalized columns in $ D $. Meanwhile, other four existing GP algorithms almost never recover the correct signal. 

\begin{figure}[htbp]
	\centering
	\subfigure[]{}
	\label{subfig1}
	\includegraphics[width=9cm, height=6.5cm]{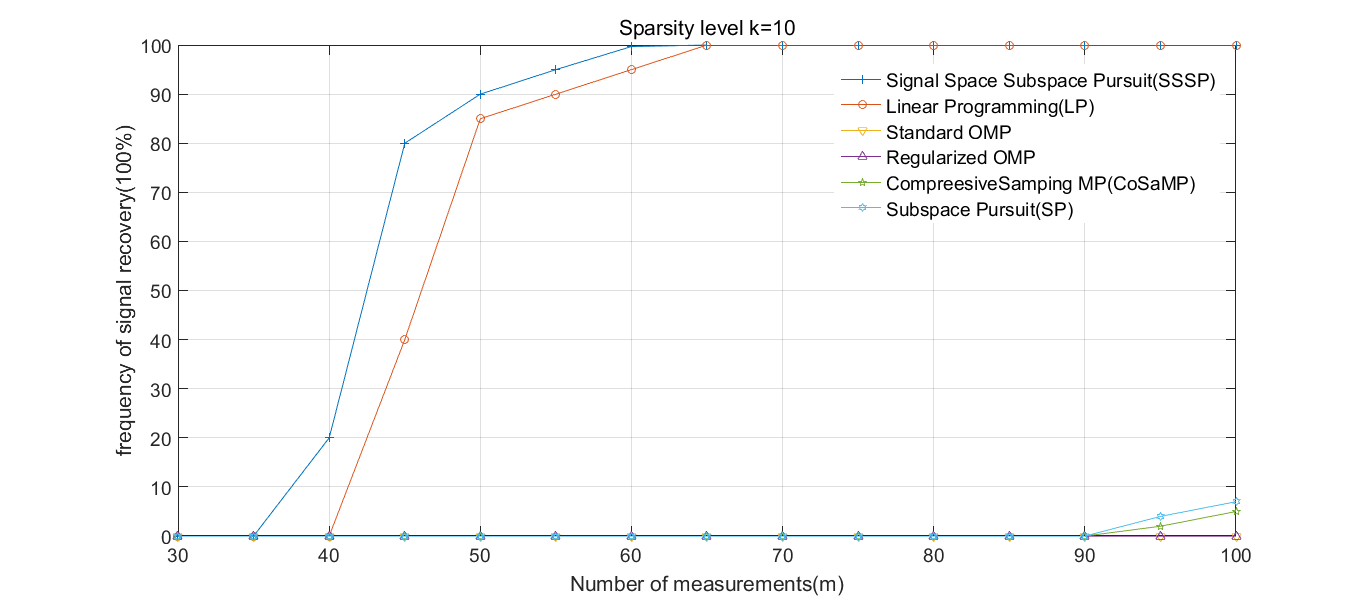}
	
	\subfigure[]{}
	\label{subfig2}
	\includegraphics[width=9cm, height=6.5cm]{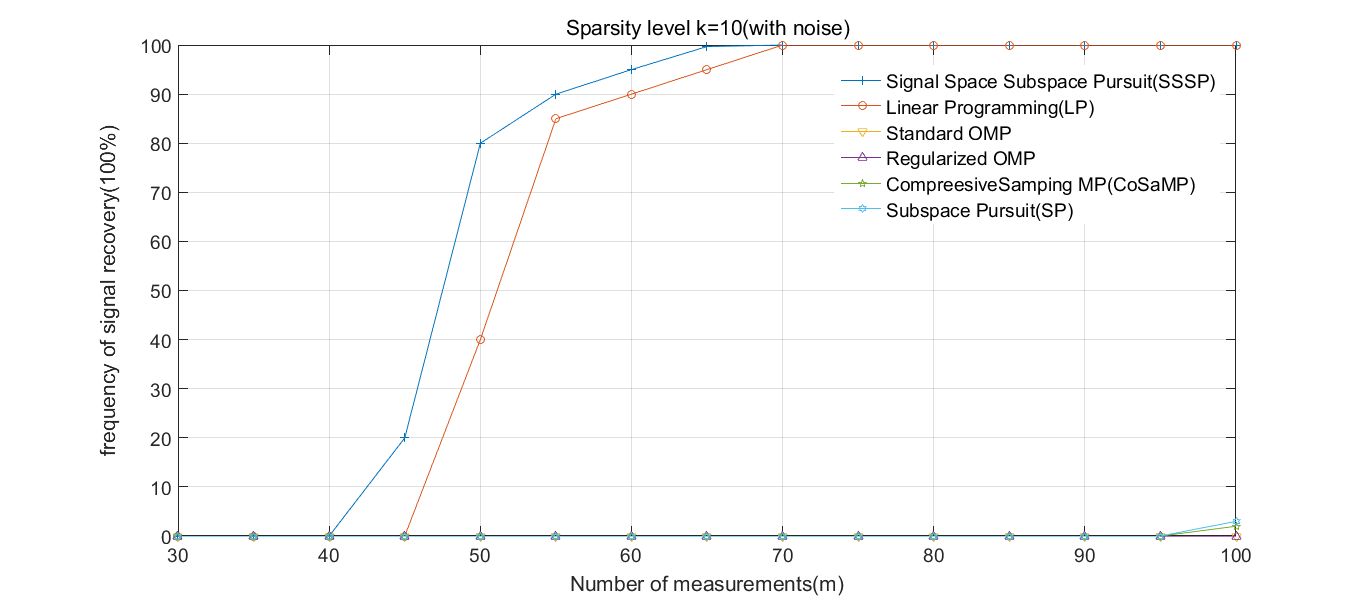}
	\caption{Performance comparison of perfect signal recovery frequency for the signals having a $ k=10 $ sparse representation in a renormalized orthogonal dictionary $ D $: (a) With noise-free measurements. (b) With noisy measurements.}
	\label{Fig.label}	
\end{figure}

\subsection{Simulation Results on Analysing the Recovery Performance of the Algorithm under an Overcomplete Dictionary}\label{formats}
In the second experiment, we check the effect of the algorithm with different support estimation techniques both for the case where the $ k $ nonzero entries of $ a $ are well separated and the case where they are clustered, and make comparisons with other four existing algorithms which includes OMP, CoSaMP, SP and LP. Note that the matrix $ D $ is a $4 \times $ overcomplete DFT dictionary. Thus, neighbouring columns are highly coherent in this dictionary. We fix the sparsity level $ k=8 $ and investigate the frequency of signal recovery as a function of the number of measurements $ m $. Simulation results are shown in Fig. 5.

As discussed in Section III-A, the main difficulty in implementing the algorithm is in calculating ${\Lambda _{opt}}(x,k)$. One such projection is required in step 1 as shown in Table I; another such projection is required in step 4.To overcome this difficulty, we apply some classical CS algorithms like OMP, SP, CoSaMP and LP to calculate the near-optimal supports ${S_D}(x,k)$. For short notation, we label 'SSSP(OMP)' when OMP is used for calculating ${S_D}(x,k)$, label 'SSSP(SP)' when SP is used for calculating ${S_D}(x,k)$, and so forth.

As can be seen in Fig. 5(a), we compare the performance of eight different algorithms for the case where the nonzero entries of $ a $ are well separated. Fig. 5(a) shows that SSSP(LP) performs better than other algorithms when using a classical algorithm like LP for the near-optimal projection ${S_D}(x,k)$. This is because that LP is available for finding ${\Lambda _{opt}}(x,k)$ exactly when $x = {P_{{\Lambda _{opt}}(x,k)}}x$ and nonzero entries of ${\Lambda _{opt}}(x,k)$ are sufficiently well separated. Also, Fig. 5(a) shows that OMP, CoSaMP, and SP are not efficient algorithms for signal recovery in this case because the sensing matrix $ A $ and the overcomplete dictionary $ D $ are highly coherent which indicates that the combined matrix $ AD $ can't satisfy the condition of the RIP.

As can be seen in Fig. 5(b), we compare the performance of eight different algorithms for the case where the nonzero entries of $ a $ are clustered. Figure. 5(b) shows that SSSP(CoSaMP) performs better than other algorithms when using CoSaMP for the near-optimal projection ${S_D}(x,k)$. This is because that CoSaMP selects 2k largest nonzero entries during each iteration and then has little effect on the coherence of neighboring active columns in $ D $. Also, Fig. 5(b) shows that SSSP(OMP) and OMP always fail with the increase of $ m $ in this case because OMP is designed to select one index at each iteration which indicates that it is not effective for recovering the correct support and will be as affected by the high coherence between close atoms in the cluster and around it. It can be seen from Fig. 5 that the algorithm yields accurate recovery whereas LP and OMP do not perform well at all when the support of $ x $ is clustered together and the exact opposite behavior is seen when the support has enough separation. Generally speaking, the algorithm variants outperform the corresponding classical CS algorithm.

\begin{figure}[htbp]
	\centering
	\subfigure[]{}
	\label{subfig1}
	\includegraphics[width=9cm, height=6.5cm]{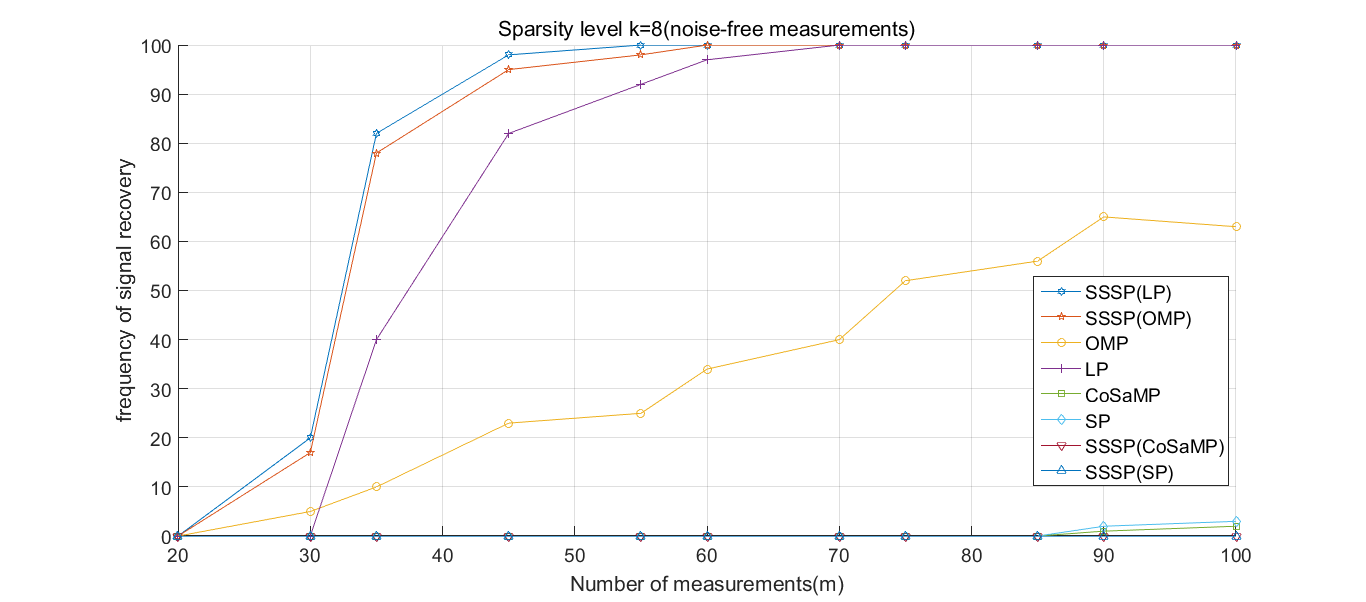}
	
	\subfigure[]{}
	\label{subfig2}
	\includegraphics[width=9cm, height=6.5cm]{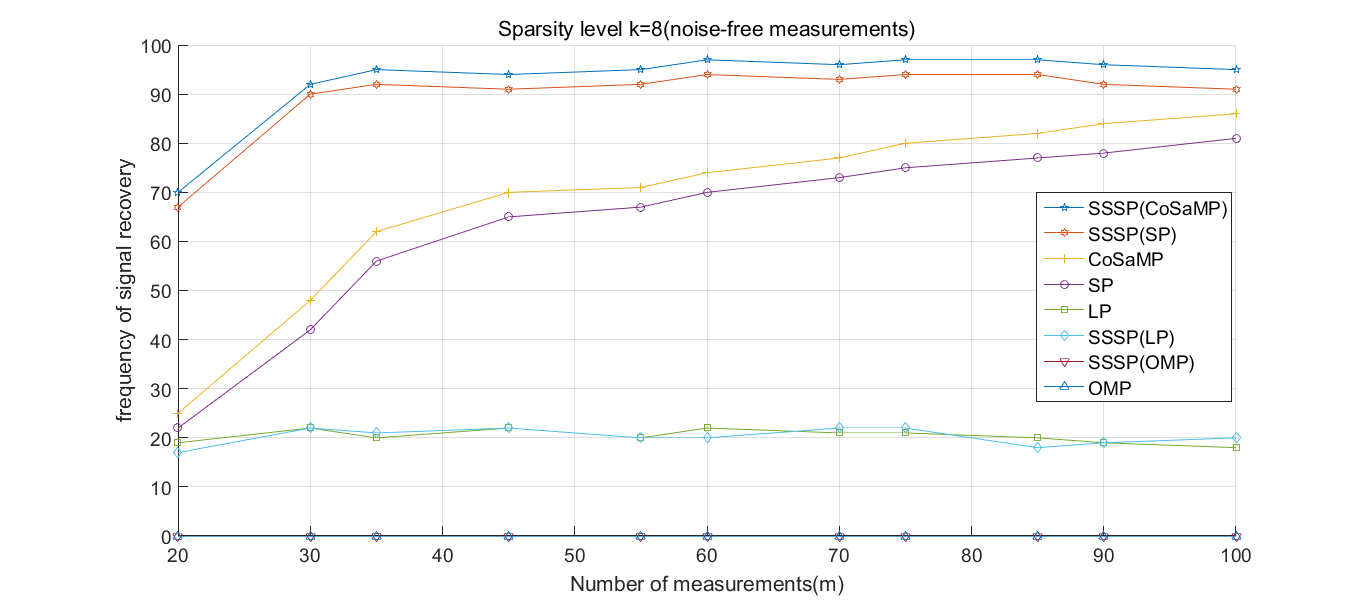}
	\caption{Frequency of signal recovery out of 1000 trials for different SSSP variants when the nonzero entries in $ a $ are well separated (a) and when the nonzero entries in $ a $ are clustered (b). Here, $ k=8 $, $ n=256 $, $ d=1024 $, the dictionary $D \in {R^{n \times d}}$ is a $4 \times $ overcomplete DFT and $A \in {R^{m \times n}}$ is a Gaussian matrix.}
	\label{Fig.label}	
\end{figure}

\section{Conclusion}
In this paper, we present support estimation techniques for a greedy sparse-dictionary signal recovery algorithm. Using this method, we propose the signal space subspace pursuit algorithm based on signal space method and establish theoretical signal recovery guarantees. We observe that the accuracy of the algorithm in this setting depends on the signal structure, even though their conventional recovery guarantees are independent of the signal structure. We analyze the behavior of the signal space method when the dictionary is highly overcomplete and thus does not satisfy typical conditions like the RIP or incoherence. Under specific assumptions on the signal structure, we demonstrate that the signal space method is used to optimally approximate projections. Thus, our analysis provides theoretical backing to explain the observed phenomena. According to the simulation results and through comparison with several other commonly used algorithms, in the noise-free and noisy cases, the algorithm achieves outstanding recovery performance. 

\section*{APPENDIX}
\subsection{Proof of condition of the Definition 3: if the dictionary is orthonormal, the value of the localization factor is one}
To complete the proof, we introduce the following Theorem.

$\bm{ Theorem }$ $\bm{ A.1 }$$\bm{ : }$ Suppose that $x \in \sum {_k} $, then 
\begin{equation}\label{eq}
	{{{{\left\| x \right\|}_1}} \over {\sqrt k }} \le {\left\| x \right\|_2} \le \sqrt k {\left\| x \right\|_\infty }.
\end{equation}

\textit{Proof:} For any $ x $, ${\left\| x \right\|_1} = \left| {\left\langle {x,sgu(x)} \right\rangle } \right|$. By aplying the Cauchy-Schwarz inequality we obtain ${\left\| x \right\|_1} \le {\left\| x \right\|_2}{\left\| {sgu(x)} \right\|_2}$. The lower bound follows since ${\mathop{\rm sgn}} (x)$ has $ k $ largest entries all equal to $ \pm 1$ (where $x \in \sum {_k} $) and thus the ${l_2}$-norm of ${{\mathop{\rm sgn}} (x)}$ is $\sqrt k $. The upper bound is obtained by observing that each of the $ k $ largest entries of $ x $ can be upper bounded by ${\left\| x \right\|_\infty }$.

\textit{Proof:} We now bound the right-hand of (11). Note that ${{{{\left\| x \right\|}_1}} \over {\sqrt k }} \le {\left\| x \right\|_2}$ in Theorem A.1. Then we have 
\begin{equation}\label{eq}
	\eta={{{{\left\| {D*Dx} \right\|}_1}} \over {\sqrt k }} \le {\left\| {D*Dx} \right\|_2}. 
\end{equation}

Notice that ${\left\| \eta  \right\|_2} = 1$ in this setting where $ D $ is orthonormal. Combining this result in (43), we have
\begin{equation}\label{eq}
	\eta  = 1 \le \mathop {\sup }\limits_{{{\left\| {Dx} \right\|}_2} = 1,{{\left\| x \right\|}_0} \le k} {\left\| {D*Dx} \right\|_2}.
\end{equation}

Using the Cauchy-Schwarz inequality again, we can get
\begin{equation}\label{eq}
	\mathop {\sup }\limits_{{{\left\| {Dx} \right\|}_2} = 1,{{\left\| x \right\|}_0} \le k} {\left\| {D*Dx} \right\|_2} \le \mathop {\sup }\limits_{{{\left\| {Dx} \right\|}_2} = 1,{{\left\| x \right\|}_0} \le k} {\left\| D \right\|_2}{\left\| {Dx} \right\|_2}.
\end{equation}

Combing (44) and (45) we see that 
\begin{equation}\label{eq}
	1 \le \mathop {\sup }\limits_{{{\left\| {Dx} \right\|}_2} = 1,{{\left\| x \right\|}_0} \le k} {\left\| D \right\|_2}{\left\| {Dx} \right\|_2} = \mathop {\sup }\limits_{{{\left\| x \right\|}_0} \le k} {\left\| D \right\|_2}.
\end{equation}

The last equation holds because ${\left\| {Dx} \right\|_2} = 1$. Thus, the (46) is equivalent to $1 \le {\left\| D \right\|_2}$. In particular, we need 
\begin{equation}\label{eq}
	{\left\| D \right\|_2}=1,
\end{equation}
where the equation follows the fact that $ D $ is orthonormal and hence ${\left\| D \right\|_2} = 1$ i.e, $D = I$. This completes the proof of condition of the Definition 3.

\subsection{Proof of Lemma 1}
To complete the proof, we introduce the following lemma.

$\bm{ Lemma }$ $\bm{ B.1 }$$\bm{ : }$ Let $A \in {R^{m \times n}}$ be a random matrix following any distribution satisfy the condition of (18). Given the assumptions for any given set $T$ with ${\left\| T \right\|_0} \le k$ and $\delta  \in (0,1)$, we have
\begin{equation}\label{key}
(1 - \delta ){\left\| x \right\|_2} \le {\left\| {Ax} \right\|_2} \le (1 + \delta ){\left\| x \right\|_2}~~~(x \in {X_T})
\end{equation}
with probability at least
\begin{equation}\label{key}
\ge 1 - 4({4 \over a}){e^{ - {c_0}(\varepsilon )m}}.
\end{equation}
where ${X_T}$ is the set of all vectors in ${R^n}$ indexed by $ T $.

Proof: Note that ${\left\| x \right\|_2} = 1$ in this case. Thus $(1 - \delta ) \le {\left\| {Ax} \right\|_2} \le (1 + \delta )$. Assume that all the vectors $ q $ are normalized, i.e. ${\left\| q \right\|_2} = 1$ for a finite set of points ${Q_T}$ with ${Q_T} \subseteq {X_T}$. Then, we have 
\begin{equation}\label{key}
\mathop {\min }\limits_{q \in {Q_T}} {\left\| {x - q} \right\|_2} \le {\delta  \over 4}~~~({\rm with}~{\left\| {{Q_T}} \right\|_0} \le {4 \over a}).
\end{equation} 

Applying (18) for the set of points with the parameter $\varepsilon  = {\delta  \over 2}$ and the probability exceeding the right side of (55) result in 
\begin{equation}\label{key}
(1 - {\delta  \over 2})\left\| q \right\|_2^2 \le \left\| {Aq} \right\|_2^2 \le (1 + {\delta  \over 2})\left\| q \right\|_2^2~~~(q \in {Q_T}).
\end{equation}

To simplify the derivation, notice that (51) can be trivially represented without the quadratic constraint on $\left\| {Aq} \right\|_2^2$ and $\left\| q \right\|_2^2$. The inequality (51) is equivalent to requiring 
\begin{equation}\label{key}
(1 - {\delta  \over 2}){\left\| q \right\|_2} \le {\left\| {Aq} \right\|_2} \le (1 + {\delta  \over 2}){\left\| q \right\|_2}~~(q \in {Q_T}). 
\end{equation}

Since ${\delta _{\min }}$ is the smallest number, thus, we have
\begin{equation}\label{key}
{\left\| {Ax} \right\|_2} \le (1 + {\delta _{\min }}){\left\| x \right\|_2}~~~(x \in {X_T})
\end{equation}

The assumption that ${\delta _{\min }}$ is the smallest number implies that ${\delta _{\min }} \le \delta $. Recall that the vectors $ x $ are normalized, i.e. ${\left\| x \right\|_2} = 1$. For a given set point $q \in {Q_T}$, the inequality (4) holds if the following inequality holds
\begin{equation}\label{key}
{\left\| {x - q} \right\|_2} \le {\delta  \over 4}.
\end{equation}

Combining (53) and (54), we have
\begin{equation}\label{key}
{\left\| {Ax} \right\|_2} \le {\left\| {Aq} \right\|_2} + {\left\| {A(x - q)} \right\|_2} \le 1 + {\delta  \over 2} + (1 + {\delta  \over 2}){\delta  \over 4}.
\end{equation} 

Because ${\delta _{\min }}$ is the smallest number for which (53) holds, the inequality (55) satisfy the following condition
\begin{equation}\label{key}
{\delta _{\min }} \le {3 \over 4}\delta (1 - {\delta  \over 4}) \le \delta.
\end{equation}

Thus, we complete the proof of the upper bound of (51). Similarly, according to the definition of ${\delta _{\min }}$, we derive the lower bound of (51)
\begin{equation}\label{key}
{\left\| {Ax} \right\|_2} \ge {\left\| {Ax} \right\|_2} - {\left\| {A(x - q)} \right\|_2} \ge 1 - {\delta  \over 2} - (1 + \delta ){\delta  \over 4} \ge 1 - \delta.
\end{equation}

Proof: Assume that there exists $C_n^k \le {({{42} \over \delta })^{2k}}$ such subspaces. The lemma B.1 shows that with probability at least
\begin{equation}\label{key}
\le 4{({{42} \over \delta })^{2k}}({4 \over a}){e^{ - {c_0}({2 \over \delta })m}}.  
\end{equation}

If $k \le {c_1}m/\log (n/k)$, then
\begin{equation}\label{key}
4{e^{ - {c_0}({\delta  \over 2})m + 2k\log ({{42} \over \delta }) + \log ({4 \over a})}} \le 4{e^{ - {c_c}m}},
\end{equation}
where both ${c_1}$ and ${c_1}$ are positive constants. The next step is to simply the both sides of (59) by leaving out the denominator exponential term $4e$ such that 
\begin{equation}\label{key}
\begin{split}
&{c_2} \le {c_0}({\delta  \over 2})m - {{2k} \over m}(\log ({{42} \over \delta }) + {1 \over k}\log ({4 \over a})) \\
&~~~\le {c_0}({\delta  \over 2})m - {c_1}({{2\log (42/\delta )} \over {\log (n/k)}} + {{2\log (a/4)} \over {\log (n/k)}}) \\
&~~~\le {c_0}({\delta  \over 2})m - {c_1}({{2(\log (42/\delta ) + \log {{(a/4)}^k})} \over {\log (n/k)}}) \\
\end{split}
\end{equation}

It is sufficient to choose ${c_2} > 0$ if ${c_1}$ is enough small. This completes the proof of Lemma 1.

\subsection{Proof of Lemma 3}
\textit{Proof:} Recall that when the number of iterations ${l+1} = \left\lceil {{{\log ({{\left\| x \right\|}_2}/{{\left\| e \right\|}_2})} \over {\log (1/{C_1})}}} \right\rceil $ holds, it can be derived that
\begin{equation}\label{eq}
\begin{split}
&{\left\| {x - {x^{l + 1}}} \right\|_2} \le C_1^{l + 1}{\left\| {x - {x^0}} \right\|_2} \\
& + (1 + p + {p^2} + ... + {p^{{l}}}){C_2}{\left\| e \right\|_2}. \\
\end{split} 
\end{equation} 

To obtain the second bound in Lemma 3, we simply solve the error recursion and note that 
\begin{equation}\label{eq}
	(1 + p + {p^2} + ... + {p^{{l}}}){C_2}{\left\| e \right\|_2} \le (1 + {{1 - {C_1}^{{l} + 1}} \over {1 - {C_1}}}){C_2}{\left\| e \right\|_2}.
\end{equation}

Combining (61) and (62) we see that
\begin{equation}\label{eq}
	\begin{split}
		&{\left\| {x - {x^{l + 1}}} \right\|_2} \le C_1^l{\left\| {x - {x^0}} \right\|_2} + (1 + {{1 - C_1^{l + 1}} \over {1 - {C_1}}}){C_2}{\left\| e \right\|_2} \\
		& \le (1 + {{1 - C_1^{{l} + 1}} \over {1 - {C_1}}}){C_2}{\left\| e \right\|_2}. \\
	\end{split}
\end{equation}

It follows that after finite iterations, the upper bound of (63) closely depend on the last inequality due to the equation of the geometric series, the choice of ${l_{it}}$, and the fact that ${x^0} = 0$. This completes the proof of Lemma 3.

\subsection{Proof of Lemma 5}
Recall that (29) in general CS model (30) is equivalent to requiring 
\begin{equation}\label{eq}
	\begin{split}
		&{\left\| {{x_k} - {x^{l + 1}}} \right\|_2} = {\left\| {D{a_k} - {x^{l + 1}}} \right\|_2} \\
		& \le 0.5{\left\| {D{a_k} - {x^{l + 1}}} \right\|_2} + 7.5{\left\| {A(x - D{a_k}) + e} \right\|_2} \\
		& \le 0.5{\left\| {D{a_k} - {x^{l + 1}}} \right\|_2} + 7.5{\left\| {A(x - D{a_k})} \right\|_2} + 7.5{\left\| e \right\|_2}. \\
	\end{split}
\end{equation}

Using the triangle inequality, we can get
\begin{equation}\label{eq}
	\begin{split}
		&{\left\| {x - {x^{l + 1}}} \right\|_2} = {\left\| {x - D{a_k} + D{a_k} - {x^{l + 1}}} \right\|_2} \\
		& \le {\left\| {x - D{a_k}} \right\|_2} + {\left\| {D{a_k} - {x^{l + 1}}} \right\|_2}. \\
	\end{split}
\end{equation}

Combing this results with (64), we obtain 
\begin{equation}\label{eq}
	\begin{split}
		&{\left\| {x - {x^{l + 1}}} \right\|_2} \le 0.5{\left\| {D{a_k} - {x^l}} \right\|_2} + {\left\| {x - D{a_k}} \right\|_2} \\
		& + 7.5{\left\| {A(x - D{a_k}} \right\|_2} + 7.5{\left\| e \right\|_2}. \\
	\end{split}
\end{equation}

Note that $ k $ contains the indices of the $ k $ largest entries in $ x $. Thus, ${x_k}$ is a best k-sparse approximate to $ x $, i.e. ${\left\| {D{a_k}} \right\|_2} \le {\left\| x \right\|_2}$. Using this to bound the right side of (66) yields 
\begin{equation}\label{eq}
	\begin{split}
		&{\left\| {x - {x^{l + 1}}} \right\|_2} \le 0.5{\left\| {x - {x^l}} \right\|_2} + {\left\| {x - D{a_k}} \right\|_2} \\
		& + 7.5{\left\| {A(x - D{a_k}} \right\|_2} + 7.5{\left\| e \right\|_2}. \\
	\end{split}
\end{equation} 

Repeat the same steps above, similarly, we can derive
\begin{equation}\label{eq}
	\begin{split}
		&{\left\| {x - {x^l}} \right\|_2} \le {2^{ - l}}{\left\| {D{a_k}} \right\|_2} + {\left\| {x - D{a_k}} \right\|_2} \\
		& + 15{\left\| {A(x - D{a_k})} \right\|_2} + 15{\left\| e \right\|_2}. \\
	\end{split}
\end{equation}

This completes the proof of Lemma 5

\subsection{Proof of Theorem 4}
To complete the proof, we introduce the following Lemma.

$\bm{ Lemma }$ $\bm{ E.1 }$$\bm{ : }$ Let ${\Lambda _0}$ be an arbitrary subset of $\left\{ {1,2,...,n} \right\}$ such that $\left| {{\Lambda _0}} \right| \le k$. For any signal $x \in {R^n}$, we define $ {\Lambda _1}$ as the index set corresponding to the $ k $ largest entries of ${x_{\Lambda _0^c}}$ (in absolute value), ${\Lambda _2}$ as the index set corresponding to the next $ k $ largest entries, and so on. Then
\begin{equation}\label{eq}
	\sum\limits_{i \ge 2}^n {{{\left\| {{x_{{\Lambda _i}}}} \right\|}_2}}  \le {{{{\left\| {{x_{\Lambda _0^c}}} \right\|}_1}} \over {\sqrt k }}.
\end{equation}

\textit{Proof:} We begin by observing that for $i \ge 2$,   
\begin{equation}\label{eq}
	{\left\| {{x_{{\Lambda _i}}}} \right\|_\infty } \le {{{{\left\| {{x_{{\Lambda _{i - 1}}}}} \right\|}_1}} \over {\sqrt k }},
\end{equation}   
since the ${\Lambda _i}$ sort $ x $ to have decreasing magnitude. Recall that when (42) still holds, we can derive
\begin{equation}\label{eq}
	\sum\limits_{i \ge 2}^n {{{\left\| {{x_{{\Lambda _i}}}} \right\|}_2} \le } \sqrt k \sum\limits_{i \ge 2}^n {{{\left\| {{x_{{\Lambda _i}}}} \right\|}_\infty }}  \le {1 \over {\sqrt k }}\sum\limits_{i \ge 1}^n {{{\left\| {{x_{{\Lambda _i}}}} \right\|}_1}}  = {{{{\left\| {{x_{\Lambda _0^c}}} \right\|}_1}} \over {\sqrt k }}.
\end{equation}

\textit{Proof:} We begin by partitioning the signal (vector) $ x $ into vectors $\{ {x_{{\Lambda _1}}},{x_{{\Lambda _2}}},...,{x_{{\Lambda _n}}}\} $ in decreasing order of magnitude. Subsets $\{ {\Lambda _1},{\Lambda _2},...,{\Lambda _n}\} $ with length ${\left| \Lambda  \right|_0} \le k$ are chosen such that they are all disjointed. Note that ${\left\| {Ax} \right\|_2} = {\left\| {\sum\limits_{i = 1}^n {{A_{{\Lambda _i}}}{x_{{\Lambda _i}}}} } \right\|_2}$. Combing this with the upper bound of (4), we have        
\begin{equation}\label{eq}
	\begin{split}
		&{\left\| {Ax} \right\|_2} = {\left\| {\sum\limits_{i = 1}^n {{A_{{\Lambda _i}}}{x_{{\Lambda _i}}}} } \right\|_2} \le \sum\limits_{i = 1}^n {{{\left\| {{A_{{\Lambda _i}}}{x_{{\Lambda _1}}}} \right\|}_2}} \\
		& \le \sum\limits_{i = 1}^n {\sqrt {1 + {\delta _k}} } {\left\| {{x_{{\Lambda _i}}}} \right\|_2} < \sqrt {1 + {\delta _K}} ({\left\| {{x_{{\Lambda _1}}}} \right\|_2} + \sum\limits_{i = 2}^n {{{\left\| {{x_{{\Lambda _i}}}} \right\|}_2}} ). \\
	\end{split}
\end{equation}

Combining (71) and (72), we see that
\begin{equation}\label{eq}
	\begin{split}
		&{\left\| {Ax} \right\|_2} = {\left\| {\sum\limits_{i = 1}^n {{A_{{\Lambda _i}}}{x_{{\Lambda _i}}}} } \right\|_2} \\ 
		&< \sqrt {1 + {\delta _K}} ({\left\| {{x_{{\Lambda _1}}}} \right\|_2} + {{{{\left\| {{x_{\Lambda _0^c}}} \right\|}_1}} \over {\sqrt k }}) \\
		& \le \sqrt {1 + {\delta _K}} ({\left\| {{x_{{\Lambda _1}}}} \right\|_2} + {{{{\left\| {{x_{{\Lambda _1}}}} \right\|}_1}} \over {\sqrt k }}). \\
	\end{split}
\end{equation}

Note that ${\Lambda _i}$ contains the indices of the ${\left\| {{\Lambda _i}} \right\|_0} \le k$ largest entries in ${x_{{\Lambda _1}}}$. Thus, maybe ${x_k}$ is a  best k-sparse approximate to $ x $, i.e. ${\left\| {{x_{{\Lambda _1}}}} \right\|_2} \le {\left\| x \right\|_2}$. Using this to bound the right side of the last inequality (73) yields
\begin{equation}\label{eq}
	\sqrt {1 + {\delta _K}} ({\left\| {{x_{{\Lambda _1}}}} \right\|_2} + {{{{\left\| {{x_{{\Lambda _1}}}} \right\|}_1}} \over {\sqrt k }}) \le \sqrt {1 + {\delta _K}} ({\left\| x \right\|_2} + {{{{\left\| x \right\|}_1}} \over {\sqrt k }}).
\end{equation}

Combining the above two inequalities yields (32). This completes the proof of Theorem 4.  

\subsection{Proof of Theorem 5}
To complete the proof, we introduce the following Lemma.

$\bm{ Lemma }$ $\bm{ F.1 }$$\bm{ : }$ Let $ x $ be an arbitrary signal in ${R^n}$. The measurements with noise perturbation $y = Ax + e$ can also be denoted as $y = A{x_k} + \hat e$ where
\begin{equation}\label{eq}
	{\left\| {\hat e} \right\|_2} \le 1.14({\left\| {x - {x_k}} \right\|_2} + {{{{\left\| {x - {x_k}} \right\|}_1}} \over {\sqrt k }}) + {\left\| e \right\|_2}.
\end{equation}

\textit{Proof:} Notice that the term ${\left\| {A(x - D{a_k})} \right\|_2} + {\left\| e \right\|_2}$ ultimately bounds the recovery error in (31). Combining this with (32), we have
\begin{equation}\label{eq}
	\begin{split}
		&{\left\| {A(x - {x_k})} \right\|_2} + {\left\| e \right\|_2} \\
		&\le \sqrt {1 + {\delta _k}} {\left\| {x - {x_k}} \right\|_2} + {\left\| e \right\|_2} \\
		& \le \sqrt {1 + {\delta _k}} ({\left\| {x - {x_k}} \right\|_2} + {{{{\left\| {x - {x_k}} \right\|}_1}} \over {\sqrt k }}) + {\left\| e \right\|_2}, \\
	\end{split}
\end{equation}  
where the last inequality (76) follows from the fact that ${\delta _k} < 1/3$ hence $\sqrt {1 + {\delta _k}}  \le 1.14$.

\textit{Proof:} Recall that (31) in Lemma 5. Thus, the recovery error for such a signal estimation can be bounded from above as
\begin{equation}\label{eq}
	\begin{split}
		&{\left\| {x - {x^{l + 1}}} \right\|_2} \le 0.5{\left\| {x - {x^l}} \right\|_2} + {\left\| {x - D{a_k}} \right\|_2} \\
		& + 7.5({\left\| {A(x - D{a_k}} \right\|_2} + {\left\| e \right\|_2}) \\
		& \le 0.5{\left\| {x - {x^l}} \right\|_2}+ 9.55{\left\| {x - D{a_k}} \right\|_2} + {{8.55} \over {\sqrt k }}{\left\| {x - D{a_k}} \right\|_1} \\
		& + 7.5{\left\| e \right\|_2} < 0.5{\left\| {x - {x^l}} \right\|_2} + 10\tilde e. \\
	\end{split}
\end{equation}

Repeat the same steps above, similarly, we have
\begin{equation}\label{eq}
	{\left\| {x - {x^l}} \right\|_2} \le {2^{ - l}}{\left\| {D{a_k}} \right\|_2} + 20\tilde e.
\end{equation}

This completes the proof of Theorem 5.

\section*{Acknowledgment}

The authors would like to thank Prof. Xu Ma and the anonymous reviewers for their insightful comments and constructive suggestions which have greatly improved the paper. 

\bibliographystyle{IEEEtran}
\bibliography{reference}

\end{document}